\documentclass[Crown,sagev,times]{sagej}

\usepackage{moreverb,url}

\usepackage[colorlinks,bookmarksopen,bookmarksnumbered,citecolor=red,urlcolor=red]{hyperref}

\newcommand\BibTeX{{\rmfamily B\kern-.05em \textsc{i\kern-.025em b}\kern-.08em
T\kern-.1667em\lower.7ex\hbox{E}\kern-.125emX}}

\begin{document}

\runninghead{Asher et al.}



\title{Strategic Maneuver and Disruption with Reinforcement Learning Approaches for Multi-Agent Coordination}

\author{Derrik E. Asher\affilnum{1}, Anjon Basak\affilnum{2}, Rolando Fernandez\affilnum{1}, Piyush K. Sharma\affilnum{1}, Erin G. Zaroukian\affilnum{1}, Christopher D. Hsu\affilnum{1}, Michael R. Dorothy\affilnum{1}, Thomas Mahre\affilnum{3}, Gerardo Galindo\affilnum{4}, Luke Frerichs\affilnum{1}, John Rogers\affilnum{1}, John Fossaceca\affilnum{1}}

\affiliation{\affilnum{1}DEVCOM Army Research Laboratory, US,
\affilnum{2}Army Research Laboratory-Research Associateship Program, US\\
\affilnum{3}University of Colorado Boulder,
\affilnum{4}Texas A\&M – Kingsville}

\corrauth{Derrik E. Asher,
DEVCOM Army Research Laboratory,
2800 Powder Mill Rd,
Adelphi, MD 20783, US}

\email{derrik.e.asher.civ@army.mil}

\begin{abstract}
 Reinforcement learning (RL) approaches can illuminate emergent behaviors that facilitate coordination across teams of agents as part of a multi-agent system (MAS), which can provide windows of opportunity in various military tasks. Technologically advancing adversaries pose substantial risks to a friendly nation's interests and resources. Superior resources alone are not enough to defeat adversaries in modern complex environments because adversaries create standoff in multiple domains against predictable military doctrine-based maneuvers. Therefore, as part of a defense strategy, friendly forces must use strategic maneuvers and disruption to gain superiority in complex multi-faceted domains such as multi-domain operations (MDO). One promising avenue for implementing strategic maneuver and disruption to gain superiority over adversaries is through coordination of MAS in future military operations. In this paper, we present overviews of prominent works in the RL domain with their strengths and weaknesses for overcoming the challenges associated with performing autonomous strategic maneuver and disruption in military contexts.
\end{abstract}


\maketitle

\section{Introduction}

In simple terms, strategic maneuver can be interpreted as a set of agents coordinating their actions to achieve a common goal by overcoming an adversary. Disruption, which is a special case of strategic maneuver, can be represented as the inhibition of an adversary's coordinated strategic maneuver. Therefore, the use of the terms strategic maneuver and disruption imply that there exists at least 2 opposing or adversarial sides that are in a dynamic struggle to gain superiority over each other by limiting, inhibiting or otherwise disrupting their opponent's coordination or tactics, and imposing their own coordinated tactics.

The nascent surge in the military modernization is motivated by the threat adversaries pose to a friendly nation in multiple domains (e.g., land, sea, air, cyber, electromagnetic, and space) \cite{sharma2020iot, sharma2020unknown,sharma2020mda}, which threatens national interest in ways beyond conventional warfare. It is expected that future battles will be fought in these complex multi-domain environments \cite{sharma2018image,sharma2019iop,sharma2019milcom}, where artificial intelligence (AI) will guide the tactics, techniques, and procedures (TTPs) of robotic agents working alongside human Soldiers. Such robots will aggregate to form intelligent multi-agent teams coordinating efficiently with human Soldiers to complete the Mission. 

The US Combat Capabilities Development Command (DEVCOM) Army Research Laboratory's (ARL) Essential Research Programs (ERPs) constitute a specific programmatic path for developing and implementing intelligent multi-agent systems (MAS). Such Army programs provide US defense operations with answers to critical research questions which converge to support the Army Futures Command (AFC) modernization efforts. Artificial Intelligence for Autonomous Maneuver and Mobility (AIMM)~\cite{aimm} and Emerging Overmatch Technologies (EOT) are example ERPs that explicitly focus on enabling the Next-Generation Combat Vehicles (NGCVs) with autonomous sensing, learning, reasoning, planning, and maneuvering capabilities. These future autonomous systems will predict and plan in collaboration with human agents and provide support to the Soldier by autonomously maneuvering in the battlefield. This paper focuses on the autonomous collaboration that needs to take place in order to have multi-agent systems (i.e., human, agent, or human and agent) succeed in future military operations.

Integrated and coordinated MAS will require technological advancements focused on collaborative strategic maneuvers beyond our current capabilities to effectively deal with equivalently (peer or near-peer) equipped adversaries. One immediate challenge is to develop teams of agents which can work autonomously and intelligently in a well-coordinated way. This demands that agents observe, orient, decide and act (OODA-Loop) alongside human Soldiers in mission critical tasks. While novel efforts are being contributed to our understanding of \emph{intelligence} in multi-agent paradigms, its current interpretation is not well-defined \cite{sharma2021survey,news2021}. Recent literature suggests that \emph{Reinforcement Learning (RL)} based approaches may provide a viable path towards such technological advances, evidenced by a substantial body of work that will be introduced in this paper. 

In this article, we set out to explore algorithmic approaches in the RL domain and their potential application in military environments; specifically on coordination through strategic team maneuver and adversarial coordination disruption in MDO. The minimization, limitation, or complete inhibition of coordination in adversarial multi-agent behaviors is a means of exploring and implementing disruption for strategic maneuvers. This avenue of overall reducing the coordination in adversarial forces will be referred to in this article as disruption. Moreover, collaborative strategic maneuvers and disruption can be learned by various RL approaches to inform a defense force of potential avenues for creating windows of opportunity or superiority.

To achieve MAS coordination through strategic maneuver and disruption with RL approaches in simulation environments, we first introduce some of the most prominent RL implementations of recent years. These recent advancements in the RL domain (e.g. alphago~\cite{silver2016mastering}) have facilitated the use of more complex multi-agent reinforcement learning (MARL) algorithmic approaches towards eventual real-world application. Further, there have been some frameworks to implement multi-agent coordination in recent years ~\cite{pierpaoli2019inferring,barton2018reinforcement,barton2018coordination,caylor2019classification,rodriguez2020measuring}. Together, these efforts may provide a path towards developing and implementing multi-agent coordination for strategic maneuver in multi-robot systems designed for the future battlefield.

In the following sections, 1) the challenges associated with utilizing RL approaches, 2) a taxonomy of RL, 3) a review of recent RL algorithms, and 4) a specific military scenario for application of RL techniques are presented. In general, these RL  approaches are shown to align with current research and development programs at ARL.

\section{Challenges}

Coordination is typically ill-defined in multi-agent tasks and is often used to indicate that a group of agents has performed successfully in some cooperative task domain. In prior work, various novel methods were developed and employed to measure the interdependence between agent actions while performing cooperative tasks, to confirm that these agents had in fact learned to coordinate ~\cite{asher2019multi,asher2019effect,asher2020multi,barton2018measuring,fernandez2021multi,fernandezemergent,zaroukian2019algorithmically,zaroukian2021emergent}. The confirmation of coordination is a precursor to establishing that a MAS is capable of working with its partners instead of simply taking actions that result in some measure of optimality. Whereas optimal behavior may be desirable in some circumstances, an agent that is simply acting optimal may result in catastrophic losses on a battlefield if the Mission has changed in some unforeseen way. Therefore, it is critical that MAS for future defense operations have the capability to explicitly coordinate.

For the remainder of this section, some of the challenges associated with developing MAS capable of strategic maneuver and disruption are described, where timescales, capabilities, and local goals may be vastly different (i.e., MDO), but some level of coordination is required. Further, it is assumed that a greater degree of flexible coordination can result in improved (e.g., faster, less losses,  non-intuitive strategies, effectively operate with changing capabilities/team composition, etc.) Mission execution.

As an environment changes in response to a dynamic battlefield, the 2 (at least) adversarial sides  may need to re-use plans and predictions in order to either 1) keep up with, or 2) stay ahead of, the planning and predictions of an opponent. An RL-trained MAS may be able to learn this dynamic planning and prediction cycle. Alternatively, this can be achieved if the learning agents build an appropriate model of their opponent's coordinated actions and then take actions to disrupt that coordination.  

In an ideal case, an algorithm (or algorithms) selected to guide behavior of a MAS would learn how to deal with changes in, the environment, adversary tactics and capabilities, own capabilities (gain new capabilities or lose previous capabilities), team composition (e.g., swapping of team mates), and mission objectives. However, current state-of-the-art (sota) algorithmic approaches are limited by experiences (i.e., MARL and RL approaches)~\cite{nguyen2020deep}. Moreover, team capabilities and composition are typically fixed in most simulations and therefore, do not provide sufficient data for an algorithm to operate on. Therefore, in selecting an algorithm to guide the behavior of a MAS intended for producing strategic maneuver and disruption for military defense applications, it is critical that novel events, behaviors, assets, and entities be considered.

In summary, current algorithmic approaches fall short of the required capability in a complex military defense MDO environment. Current shortcomings can be divided into three categories: 1) data requirements, where either data is limited due to the novelty of a  situation, a data set is insufficient to produce accurate predictions, or the data is polluted in some way (e.g., noisy, dirty, or adversarial alteration), 2) limited computational resources, and 3) algorithms do not generalize to situations other than what was encountered during training (e.g, different goals, altered capabilities, or modified team composition) resulting in a narrow or brittle MAS solution.

In this work, we narrow our focus to RL algorithmic approaches that can guide a MAS to overcome the aforementioned challenges for strategic maneuver and disruption in military defense MDO. Technically, RL is a branch of machine learning (ML) algorithms that go beyond building accurate predictions from data, by demonstrating learning with the production of actions in an environment, which can also be considered a form of decision-making, but more accurately described as strategic action selection. 

RL agents learn (or are trained) based on a reward function that ultimately determines which is the best action for an agent to select given the current state of that agent (i.e., current situation). For example, an RL agent can interact with an environment to produce experiences that are tied to rewards (based on a reward function), which will result in a learned policy (i.e., a series of state-action pairs). However, we will discuss in later sections that RL algorithms are not yet mature enough to overcome the challenges that would result in human-like adaptability for intelligent decision-making in novel situations or environments. Although RL algorithms have their shortcomings~\cite{dulac2019challenges}, they appear to be the most promising avenue moving forward towards achieving successful strategic maneuver and disruption in military defense MDO environments. 

In the next section, we address RL shortcomings in greater detail to illuminate how overcoming such issues may produce solutions for military defense MDO environments. To do so, an existing taxonomy of RL algorithms is introduced. This effort should provide a better insight into promising RL techniques that may help determine viable avenues for eventual application in US defense MDO environments.    

\section{Reinforcement Learning Taxonomy}
According to Barto and Sutton~\cite{sutton2018reinforcement}, reinforcement learning is learning what to do -- how to map situations to actions -- so as to maximize a numerical reward signal. This general description of RL indicates that agents select actions in some state-space that maximize a reward signal, resulting in a policy, which can be interpreted as a series of state-action pairs (i.e., given a state, an action was learned) and these learned policies determine the trajectory of future actions selected by an agent.

An RL algorithm has one or more of three major components: 1) value function, 2) policy, and 3) model. The value function represents how good it is to be in a state, which can be used implicitly to derive a policy. The policy describes the agent behavior; what actions an agent should take from any given  state. The model represents the environment an RL agent interacts with which can be represented as a Markov Decision Process (MDP)~\cite{howard1960dynamic}. RL algorithms can be taxonomized in three categories based on their use of value functions or policies to guide learning and actions: 

\begin{itemize}
    \item Value based: value function  (implicit policy)
    \item Policy based: policy
    \item Actor Critic: value function and explicit policy
\end{itemize}

However, the most prominent difference between RL algorithms is whether a model of the environment is utilized (termed model-based or model-free). So, the aforementioned taxonomy can be refined further:

\begin{itemize}
    \item Model Free RL
    \begin{itemize}
        \item No environmental model is utilized
        \item Value-based and/or Policy-based  
    \end{itemize}
    \item Model-based RL
    \begin{itemize}
        \item Use of an environmental model
        \item Value-based and/or Policy-based  
    \end{itemize}
\end{itemize}
Finally, combined model-free and model-based RL algorithms fuse together these approaches to get the best of both worlds. In the next two sections, basic RL algorithms are described. Further, technical details are presented for some popular RL algorithms to illuminate their drawbacks, strengths, and  potential to be adapted for strategic maneuver and disruption in military domains. 

\section{Background}

\subsection{Dynamic Programming and Markov Decision Processes}
Dynamic Programming (DP) deals with problems that can be solved in a step-wise approach to obtain an optimal policy through iterations. DP can be used for prediction to compute the value function given a policy. However, DP also assumes that the MDP is known. Therefore, DP can also be used for optimal control: finding the optimal policy by evaluating over several different policies.

There are at least two methods for developing a policy in RL, these are policy iteration and value iteration. In policy iteration~\cite{sutton2018reinforcement} the Bellman expectation equation is used iteratively to evaluate a policy and then an   iterative greedy policy improvement method is used to get to an optimal policy. In value iteration, the control problem is solved with the Bellman optimality equation followed by a single optimal policy extraction.  

In RL the terms ``model-based'' and ``model-free'' do not refer to the use of a neural network or  statistical learning model to predict values, or a next state. Instead, these terms refer strictly to the agent's use of state transition probability to solve the problem of state-action-selection in an environment. An agent can simply use a model (model-based) to predict the next state and the current state's value. These models of the environment can be provided entirely outside of the learning agent (e.g., by an outside observer that understands the rules of the task). Or the model can be learned by an agent (model-free) through trial and error experiences, where the agent in some state selects an action, and learns the reward associated with that state-action pair.

If an MDP is unknown, a model-free approach is needed to find a policy. In model-free approaches, an RL agent interacts with the an environment to gain experience and learn to take rewarding actions. Further, In model-free prediction the value function is computed given a policy (i.e.,the policy is evaluated). The challenge with a model-free approach is that the value function must be learned with no knowledge of how the world (environment) works. 

\subsection{Monte Carlo Temporal Difference and Q-Learning}
Monte Carlo (MC) RL methods~\cite{hammersley2013monte} solve the prediction problem from direct experiences of an episodic (restarts after some duration) environment. In MC learning, an agent selects actions with or without receiving feedback (i.e., reward signal) until the end of an episode \footnote{An episode is an amount of time (either a preset duration or the episode is stopped because a predetermined condition was met) that an agent has to explore the environment through trial and error actions (or state-action pairs)}. Once an episode (or batch of episodes) has completed, the value or return for that series of state-action pairs \footnote{Each forward step through time (i.e., time step) produces a state-action pair} is computed as the value function. On the other hand, temporal difference (TD) learning~\cite{sutton2018reinforcement} utilizes incomplete returns to estimate the value of a state. This is a form of bootstrapping, which provides the agent with an estimate of the predicted reward all the way to the end of an episode, although the agent had only taken one step. In contrast, TD($\lambda$) learning agents can combine MC and TD learning methodologies and take a variable number of steps to estimate the value function.

RL model-free control methods can be divided into on-policy and off-policy learning. On-policy indicates that an agent learns from experiences that are gained through interactions with an environment. This implies that on-policy techniques evaluate a policy by actively following a current policy, then iteratively updating the policy during experience collection. In stark contrast, off-policy  methods improve a policy by iterating over previously collected experiences \footnote{The experiences that an off-policy method utilizes to improve the policy can come from non-policy related behaviors (e.g., demonstrations from an expert in the domain or a series of human directed state-action pairs)}. Q-learning is an off-policy learning method that utilizes a separate, sometimes greedy policy method to sample and estimate the discounted return from the environment, while iteratively updating a different policy.
 
MC methods can yield high variance in the policy (i.e., the agent's learned behaviors). To solve the issue, a MC approach can be replaced with a TD learning approach to update the Q-value (action-value function) and instead, use an $\epsilon-$greedy method to improve the policy. The benefit of TD learning is that an estimated return from the end of an episode can be bootstrapped, which effectively reduces policy variance. An algorithm that utilizes this particular approach is referred to as Sarsa~\cite{sutton2018reinforcement}. 
 
Sarsa can be modified to accommodate multiple time step intervals so that a policy can be bootstrapped from any point in an episode, which is called n-step Sarsa. Further, the time steps can provide variable feedback to the policy with a weighted return across the different points, which is referred to as Sarsa($\lambda$). Instead of a forward looking, on-policy approach, a backward looking method can be constructed with eligibility traces for every state-action pair. With this backward looking method, a Q-value can be updated in proportion to the eligibility trace along with a TD-error signal. The combination of these methods provides the Sarsa algorithm an opportunity to project forward through episodic time or backwards to generate an RL policy that may be capable of learning in novel situations. However, Sarsa is inherently a single agent algorithm and may need further modifications to be adapted to MAS problem spaces, where partner or adversary actions can change the goals of the environment (i.e., non-stationarity).

\section{Scalable RL}

Scalability of a learning algorithm is one of the major concerns for military tasks in MDO, especially since  such tasks may require a large number of agents to complete an objective. In addition, military tasks can involve multiple sub-tasks each with their own sub-goals, further complexifying the scenario. In MDO, it is expected that a sub-goal consists of a myriad of complex strategic maneuvers that would require fast computation for MAS, and an optimal (or at least sufficient) strategy with the use of minimal computational resources (e.g., computing at the tactical edge). Therefore, a scalable RL algorithm must account for 1) environmental and task complexities, and 2) number of agents (partners and adversaries) so each agent can properly select actions as experiences are collected through the RL learning process. 

Environmental complexity (i.e., the size of an agent's state and action spaces) can refer to the number of states available in an environment's state space along with the number  and actions available to agents in that environment. Scalability of an RL algorithm is the capability of computing an optimal policy within reasonable time and computational power for a sufficiently complex state and action space \footnote{Sufficiently complex is used here as an arbitrary term that is typically identified post-hoc through trial and error methods (e.g., different state and action spaces}. Environmental complexity also entails the inclusion of additional agents (e.g., expanding to a MAS), where the state space is scaled up to account for the extra agents and the size of the action space is multiplied by that number of agents.  

It is not practical to tackle the scalability issue in RL by using a table for state-action pairs because continuous domains would make the table untenable, and simultaneously updating the entries of a table for all agents is  infeasible within a reasonable amount of time. Even with sufficiently large computational resources (e.g., excessive computer memory) to contain all states, learning across each state-action pair would be too slow. In contrast to utilizing a table for tracking state-action pairs, a solution is to use a non-parametric function approximator (e.g., deep neural network where the weights are the parameters) which approximates values across the entire state space. However, a function approximator must be differentiable, such that a gradient can be calculated to provide the direction of parameter adjustments. 

There are two approaches to train value function approximators: 1) incremental methods, and 2) batch methods. Incremental methods use a stochastic gradient to adjust the approximator's parameters in the direction of the gradient to minimize the error between the estimated and target values. However, the incremental approach is not sample efficient, and therefore, does not lend itself to scalability. In contrast, batch methods save the data from a set of experiences and use them to compute the error between function approximator estimation and the target value. Batch methods share commonalities with traditional supervised learning, where the outcome is known (e.g., data is labeled) and an error is calculated between the approximator's estimate value and the actual outcome value. This type of batch learning is typically referred to as experience replay. Repeating this process will lead to a least square error solution. A recent successful example of experience replay was demonstrated with deep Q-networks (DQN)~\cite{mnih2013playing} playing Atari games. Although function approximator methods have shown success in complex environments, it is unlikely that this approach alone will be sufficient to train a MAS for MDO scenarios without accounting for the inclusion of additional agents (i.e., non-stationarity or partial-observability).

Compared to value function approximation, policy learning methods rely on Policy Gradient (PG) calculations to explicitly optimize a policy rather than indirectly on a value function. PG has better convergence properties over function approximator methods \footnote{Policy convergence is an important property that implies an algorithmic approach reaches a stable level of reward and performance after a sufficient number of training iterations}. The main reason PG methods are used over value approximation methods is their ability to be effective in high dimensional and continuous action spaces (i.e., scalable in complex environments). In an MC policy gradient (e.g., REINFORCE algorithm \cite{williams1992simple}) the actual return is multiplied with a score function to get the gradient to adjust the policy parameters to obtain more reward. An MC policy gradient still has high variance and is slow to converge, since it uses the entire trajectory of an agent's state-action pairs across an episode to get the return value. An alternate solution that may surpass the shortcomings of traditional function approximator methods is to utilize Actor-Critic approaches. 

With Actor-Critic approaches, a PG equation is modified to use the value function approximation instead of using the true action-value function multiplied by the score (as is done with the REINFORCE algorithm \cite{williams1992simple}). This indicates that the actor adjusts the policy in the direction that the critic is pointing so that total cumulative rewards can be maximized. This policy evaluation step by the critic can be done by using combined value approximation methods (i.e., MC, TD(0) and TD($\lambda$)). To reduce the variance in the policy gradient, an advantage function can be used \cite{mnih2016asynchronous}. The advantage function tells us how much better one action is over another (Q-value) compared to a general state value function. This implies that the critic must estimate the Q-value. An efficient way to do this is to use TD-error, which is an unbiased sample of the advantage function where the critic approximates one set of parameters. TD($\lambda$), eligibility traces can also be used for the critic to estimate the value across different time steps. Interestingly, MC (high variance) and TD methods can be used with the actor to consider modify the policy across time steps.

Since MDO involves military tasks where an RL algorithm must have the capability to coordinate with many other agents for optimal strategic maneuvers and disruption, a MAS must be able to scale with a large number of agents and heterogeneous assets. Another important capability of an algorithm is the ability to process the vast observations of complex state-spaces (i.e., many agents) and multi-domain environments. In the next sections, we discuss the implication of using different kinds of RL algorithms in MDO for strategic maneuver and disruption.

\section{Model-Free RL}

Model-free algorithms can be divided into off-policy and on-policy algorithms where state-action space can be either continuous or discrete. In this section, the strengths and weaknesses of the model-free algorithms are discussed, along with how they might align with strategic maneuver and disruption, leading to the goals of MDO. The purpose of this analysis is to provide direction towards finding potential algorithmic approaches that may achieve strategic maneuver and disruption in MDO environments.

\subsection{Deep Q-Network}

Deep Q-Network (DQN)~\cite{mnih2013playing} is a single RL agent algorithm that was trained to play Atari 2600 games~\cite{bellemare2013arcade} where the action space was discrete and the state space was continuous. DQN uses a convolutional neural network trained with Q-learning~\cite{watkins1992q} to learn from high dimensional input (sequential images). 

The DQN algorithm is a sample efficient approach since it makes use of all the collected experiences to extract the maximum amount of information possible. DQN is robust enough to be trained using the same hyperparameters \footnote{In machine learning, a hyperparameter is a parameter whose value is used to control the learning process. By contrast, the values of other parameters (typically node weights) are derived via training.} to play six different Atari games, where the agent performed better than human experts in three of these games. 

However, a drawback with DQN is that there are no theoretical guarantees of a trained neural network achieving a stable Q-Value prediction (i.e., potentially high-variance in the trained policies across independent models). 

Given that DQN is inherently a single RL agent model, it should be insufficient for strategic maneuver and disruption in MDO. In MDO,  multi-agent RL algorithms may be more suitable due to the typical decentralization of the agents during execution time, allowing for agents to operate independent from one another. In addition, the original implementation of DQN only utilizes a sequence of four observations to learn the Q-value, which is insufficient for strategic maneuver and disruption in MDO. Strategic maneuvers of multiple assets can not be typically be captured within such short time intervals. In fact, this is the primary reason that DQN did not perform well compared to humans in three of the Atari games evaluated (i.e., Q*bert, Seaquest, and Space Invaders). However, there exist some variations of DQN to address this  and other weaknesses. 

Bootstrap DQN~\cite{osband2016deep} is one such variation that learns an ensemble of Q-Networks to improve sample efficiency and overcome the shortcomings of traditional DQN. Action elimination is another method used with DQN to tackle large action spaces~\cite{zahavy2018learn}. DQN with a type of memory (i.e., recurrent neural network) can be used to handle partial observability as well~\cite{hausknecht2015deep}. This approach is particularly useful if an agent needs to  navigate an environment for task completion. Alternatively, distributional DQN~\cite{bellemare2017distributional, dabney2018implicit} returns a distribution which can be used to evaluate policy risk and to reduce variance or noise around an optimal solution.

Although DQN and its modified variants are promising for tackling tasks more complicated than simple Atari games, the DQN method inherently lacks a multi-agent prediction mechanism to conduct coordinated tactics, which are required for strategic maneuver and disruption in MDO. Further, DQN is most often too computationally intensive to be used in militarily relevant environments. Finally, DQN algorithmic approaches lack sufficient adaptability for unseen examples (e.g., novel behaviors of partners or entities/obstacles emerge in an environment). 

\subsection{Deep Deterministic Policy Gradient}

In the real world, most regular tasks involve continuous state and action spaces. However, DQN only considers discrete state spaces and low dimensional action spaces. An alternative approach to DQN, where continuous state and action spaces are handled, is the Deep Deterministic Policy Gradient (DDPG) method. DDPG advances the progress from DQN approaches by combining value function approximation and deterministic policy gradient (DPG)~\cite{silver2014deterministic}. DDPG utilizes an actor-critic approach~\cite{heess2015learning}, which can overcome the complexities of continuous spaces.  This model-free, off-policy, prediction and control algorithm can perform physical control tasks (e.g., cart pole, dexterous manipulation, legged locomotion, or car driving). 

Another approach that uses a deep neural network is Trust region policy optimization (TRPO). This method constructs a stochastic policy directly~\cite{schulman2015gradient} without the need for an actor-critic model (not to be confused with an environment model which would make this a model-based approach). Similar to TRPO, guided policy search (GPS)~\cite{levine2016end} is void of an actor-critic model and uses trajectory guided supervised policy learning along with some additional techniques (e.g., reduction of dimension from visual features, additional information on robot configuration dynamics at the first layer of the network). As a result, GPS is data-efficient and can be adapted to DDPG if required. PILCO~\cite{deisenroth2011pilco} on the other hand learns a probabilistic model first, then finds an optimal policy. PILCO is highly data efficient in some problem domains, however, it is computationally demanding. Further,  D4PG~\cite{barth2018distributed} proposed some improvements over the DDPG algorithm: distributional critic update, distributed parallel actors, N-step returns, and prioritization of the experience replay to achieve a more stable and better solution for different category of tasks. 

From the perspective of strategic maneuver and disruption, the primary drawback of the DDPG algorithm is that it was designed as a fully decentralized single agent algorithm (i.e., independent learners). As such, the DDPG algorithm does not facilitate coordination in multi-agent scenarios.  Consequently the resulting strategic maneuvers using DDPG will not result in coordinated team behaviors. Moreover, DDPG is not equipped to handle role based tasks with multiple objectives which might be a requirement for disruption.

\subsection{Multi-Agent Deep Deterministic Policy Gradient}

RL agent interaction is crucial to AI systems for strategic maneuver and disruption where different agents may need to form teams to effectuate strategic collaboration against an adversary (strategic maneuver) or inhibit the adversary's coordination (disruption). Q-Learning and policy gradient (PG) approaches alone suffer from non-stationarity \footnote{Non-stationarity generally refers to an environment where sudden concept/goal/task drift can occur due to dynamic and unknown probability data distribution functions associated with other agents taking actions that change local goals.} and high variance respectively. To overcome these issues, the Multi-Agent Deep Deterministic Policy Gradient (MADDPG)~\cite{lowe2017multi} algorithm extends an actor-critic approach, which allows it to work for multi-agent systems by centralizing agent training. The MADDPG framework adopts a centralized critic and for training and deploys decentralized actors during test time. A critic (one for each agent) receives the policy of every agent, which allows for the development of dependent policies with potentially different reward functions (e.g., MADDPG permits training of adversarial teams with opposing reward functions). Conversely, the actors (i.e., policy networks) only have local knowledge during training and testing. The actor improves the policy iteratively (through training) in a direction consistent with the critic's evaluation.

Although, a major weakness of MADDPG is that the input to the Q-function increases with the number of agents of the environment (not scalable). This poses a problems for strategic maneuver and disruption in MDO. If agents need to be replaced, added, modified, or removed, retraining may need to take place. In strategic maneuver and disruption, agents may need to switch roles or change capabilities periodically, which poses a major challenge towards adapting MADDPG to military domains. In addition, frequent retraining would make rapid strategic maneuver unlikely. Reducing training time will reduce the computational load on the edge and make rapid strategic maneuver or disruption possible. MADDPG can not accommodate such extreme cases. For military applications, a robust model of opponents or agents is desired so that the operational period is maximized (i.e., enough time to execute strategic maneuver and disruption). 

A potential modification to MADDPG to address its scalability issues is to form clusters of agents and learn a policy for the clusters instead of each agent individually. In the case of a new event, the need for retraining can be postponed because, in theory, a cluster of agents would have a set of variable capabilities to handle dynamic situations. Further, this would avoid increasing the input space for the Q-function as agents are modified or new agents are introduced. However, the question arises; how we can decompose a task into partially independent sub-tasks with minimum degradation of an optimal group policy? 

Although, MADDPG can lead to a set of heterogeneous multi-agent policies capable of diverse tasks, this approach does not scale well beyond a dozen agents. As the number of agents grows, the variance of the policy gradient grows exponentially. Therefore, this approach is not well suited for strategic maneuver and disruption in MDO where more than forty heterogeneous agents must be accounted for in adversarial contexts. A method for overcoming this scalability issue is the Mean Field Multi-agent RL algorithm ~\cite{yang2018mean}, which computes a mean estimation for the neighborhood agents' Q-value that may result in high error margin when the nearby interaction between agents get complex. Further, the Evolutionary Population Curriculum~\cite{long2020evolutionary} (EPC)  algorithm was designed to make MADDPG scalable by combining genetic algorithmic approaches with RL. With advances upon MADDPG and the successes shown with the approach, it is conceivable that these algorithmic advances could lead to robust demonstrations of strategic maneuver and disruption within MDO in simulation experiments.

Distinct from MADDPG, the Counterfactual Multi-Agent (COMA)~\cite{foerster2017counterfactual} approach uses a single centralized critic for all agents but is designed for  discrete action spaces. COMA is more  scalable than MADDPG, but it may result in a homogeneous set of policies that could fail with sufficiently different agent capabilities, different local goals, or different reward functions. Similar to MADDPG, Minmax Multi-Agent DDPG (M3DDPG)~\cite{li2019robust} adds an improvement over the original version of MADDPG, by allowing agents to develop more robust policies against adversaries (i.e., competitive games with opposing reward structures). However, M3DDPG is still unable to handle scenarios when heterogeneous agents  are introduced into the system.

Implementing algorithms into environments with continuous state and action spaces sometimes requires utilizing common techniques to manipulate the inputs or outputs, such as discretizing the state and action spaces or converting the discrete policy output to a continuous output. One example of converting the policy output is the implementation of MADDPG in the OpenAI Multi-Agent Particle Environment. In this example, the discrete policy components are utilized to compute continuous actions. In another perspective, the multi-agent transformer soft double Q-learning algorithm~\cite{hsu2021scalable} discretizes the continuous action space into a set of velocity and angular velocity controls which can then be used in a motion model. Although these techniques permit the use of such algorithms in continuous environments, these algorithmic approaches do not train with continuous information, which could limit their efficacy in physical environments for strategic maneuver and disruption.

\subsection{Value Based}

A recent family of value-based MARL algorithms~\cite{pymarl} has proven to be quite successful in the very complex Starcraft 2 simulation environment~\cite{smac} where a centralized joint action-value $Q_{tot}$ is learned based on the agents' local $Q_a$ values. A decentralized policy is then extracted from the $Q_a$ by taking the linear argmax operator. This very simple but efficient factorization approach avoids learning the joint action-value which does not scale very well. If new agents are added or agents are replaced with new capabilities retraining still has to be done. However, it is more scalable compared to MADDPG because the individual Q-values are learned from local observation only which avoid learning the joint-action value by learning a factorized $Q_{tot}$. Still the scalability of this family of algorithms can be challenged when there are more than forty agents. To make it more scalable role based algorithm RODE~\cite{wang2020rode} has been proposed where the agents' roles are determined by clustering their actions based on their effect on the environment. The algorithm has shown very promising results for large number of agents. 

For strategic maneuver and disruption the RODE algorithm is very promising since groups of agents can be assigned to different roles, where roles can be based on their action and effects on the environment or any other fixed behaviors (for ally or even enemy). The algorithm then can be used for strategic role switching for different groups. Since the action space of different roles are restricted, the algorithm converges very quickly. This algorithm is also fit for strategic use of role based disruption which we might consider for out future work. Even though RODE is very scalable it is not clear how we can adapt it for when new agents will be added to the environment since a centralized policy needs to be learned for optimal coordination. 

In contrast to the RODE algorithm, a scalable multi-agent reinforcement learning method~\cite{hsu2021scalable} deploys an entropy-regularized off-policy method for learning a stochastic value function policy that has been experimentally shown to be able to scale to over 1000 agents. As discussed previously, scalable RL algorithms are concerned with the complexity of the environment -- the more agents in the system or team, the larger the state space. RODE is limited as it uses a centralized policy that must be retrained when more agents are introduced into the environment. The algorithm, multi-agent transformer soft double Q-learning, is a centrally trained off-policy learning algorithm, i.e. sharing a central experience replay buffer, with decentralized execution, i.e. each agent makes its own control decisions based on its local observations, not from a central controller. Due to this decentralization scheme, when  agents are added or removed from the system, the team is unaffected and continues to execute their policy.

With respect to scalability, training a large MAS (i.e., many agents in the team) is difficult, and it has been shown that even state of the art algorithms fail to learn performant policies for complex MARL tasks. Multi-agent transformer soft double Q-learning alleviates this scalability issue by utilizing a heuristic during training that allows for the policy to be trained on a smaller set of agents (e.g., 4 agents tracking 4 targets in a target tracking scenario), and the policy has been shown to work with many more agents in execution without any adaptation needed (i.e., tested and evaluated with 1000 agents). The heuristic used during training and execution allows the algorithm to address a dramatic distribution shift in the number of agents: it essentially scales down the large complex observation space at test time into something that is close to what the agent policy was originally trained for. In the military perspective, this formulation is ideal for strategic maneuver and disruption as agents in the field might be lost or gained in-situ and might have to account for additional strategic information. A flexible and scalable algorithm provides the capabilities needed to be robust in MDO.

\section{Model-Based RL}

In contrast to model-free RL approaches where agents learn a policy from a stream of experiences (i.e., no prior environmental knowledge), model-based RL approaches enable agents with an a priori functional representation of an environment (i.e., given some pre-designed model of how an environment functions). Model-based RL has a strong advantage over model-free RL by being sample efficient (i.e., less computational resources needed). This indicates that an agent needs very few samples to learn how to achieve a goal within some environment. Once the model and the cost function are known, an agent can operate optimally without further sampling (i.e., no additional real-world experiences are needed but simulated experiences are used). However, providing an appropriate environmental model can be quite difficult and in the case of strategic maneuver and disruption within MDO, perhaps impossible.

In model-based RL, a model can be learned via supervised learning. The rules of a situation can provide sufficient information for a model. Also, the model-based approach provides a clear opportunity to reason about the familiar and unfamiliar states of an environment to explore and exploit more effectively. If an environmental model is not accurate, uncertainty can be reasoned about through exploration, similar to a  model-free approach. A model-based approach for strategic maneuver and disruption in MDO can be promising when small factors can affect the value and policy drastically. 

Potential applications for model-based approaches for strategic maneuver and disruption in MDO can be explored with game-based simulations (e.g., adversarial scenarios involving turret reconnaissance, predator-prey pursuit, target acquisition, or multi-asset fire-fight games with well structured rules and win conditions) where the state-space is relatively small. In game-like cases, learning an environmental model is potentially efficacious, whereas real-world environments may have untenable environmental models. An example in literature is the alphago~\cite{silver2018general} algorithm, where agents were given a model of the game environment (the game of Go), and the model-based approach performed better than model-free RL approaches. Another disadvantage of model-based approaches is that  there are two sources of approximation error: policy (model-free) along with the environmental model (model-based).

Forward search is another example of model-based RL that focuses primarily on planning. Forward search does not solve the entire MDP. It focuses on a sub-MDP space that is simply taken from an agent's current state. The sub-MDP is solved by forwarding search approaches that select the best action via \textit{lookahead} techniques. In general, forward search algorithms construct a search tree with from the current state as a root, using a model of the environment to \textit{lookahead} to the expected reward for a series of actions towards the goal. 

Simulation-based search is a forward search technique that uses sample-based planning. The samples are generated using a model of the MDP to estimate what might happen in the future. Since the samples are generated based on a model of the environment, the samples reflect the most probable experiences rather than focusing on extraneous, unnecessary, or possibly pointless experiences through environmental exploration (as is done in model-free approaches). This approach might be useful for situation when a strategic maneuver and disruption policy need to be learned within a very short amount of time.  

An alternative to forward search methods in model-based RL, Monte Carlo (MC) or TD techniques can be used in conjunction with a model of an environment. Monte Carlo Tree Search (MCTS)~\cite{silver2016mastering, silver2017mastering, silver2018general} is a specific example of a simulation-based search technique that has proven to be successful in achieving human-like performance in games like Go. However, if we consider strategic maneuver and disruption in MDO, MCTS needs to be extended to multi-agent systems. In fact,  the multi-agent version of MCTS (i.e., MAMCTS~\cite{zerbel2019multiagent}) has shown promising performance in finding coordination strategies compared to other similar algorithms in  grid-based worlds (i.e., discrete state-spaces). Further, MCTS can operate in continuous action spaces~\cite{yee2016monte, lee2020monte}. 

MC search methods can be made more efficient with bootstrapping techniques like TD learning (one step, n-step, or TD($\lambda$)), which is referred to as TD search. Alternatively, a function approximator may be used to estimate Q-values instead of either MC or TD approaches.
 
Providing an accurate enough model for agents to explore viable behaviors that lead to successful outcomes is the core challenge using model-based RL for strategic maneuver and disruption in MDO. 
Among all the model-based RL techniques explored here, it appears that simulation-based approaches are the most promising since they have the potential to provide robust action-selection in situations where agents experience drastic changes to the environment is unseen or novel situations.  Since simulation-based search solves an MDP only from the current state, this can be highly promising for minimizing execution time while maintaining a robust policy in dynamic situations, such as strategic maneuver and disruption in MDO. 

\section{Model-Free and Model-Based RL}

Dyna architecture~\cite{sutton2018reinforcement} is an example of an algorithm that combines model-free and model-based approaches. In this architecture gathered experiences are used to learn a model of the environment, where  both model-based errors and model-free (gathered) experiences are combined to learn the value function and policy. 

For strategic maneuver and disruption in MDO, the Dyna architecture can provide a balance between long term and short term needs associated with different tasks. However, in an adversarial scenario (i.e., agents have opposing reward structures), where a value function can vary drastically, model-based approaches might be more beneficial than model-free approaches due to opposing sides adopting new exploitive tactics. Dyna-Q+ improves upon the Dyna-Q algorithm by providing reward explicitly for exploration (i.e., promoting model-free methods). Although, Dyna-Q+ has been explored minimally in multi-agent paradigms. However,  Dyna-Q+ has  potential for strategic maneuver and disruption in MDO because of its promising capability of handling unseen examples in policy formation. 

\section{MAS for Strategic Maneuver and Disruption in MDO}

\begin{figure}[t!]
    \centering
    \includegraphics[scale=0.4]{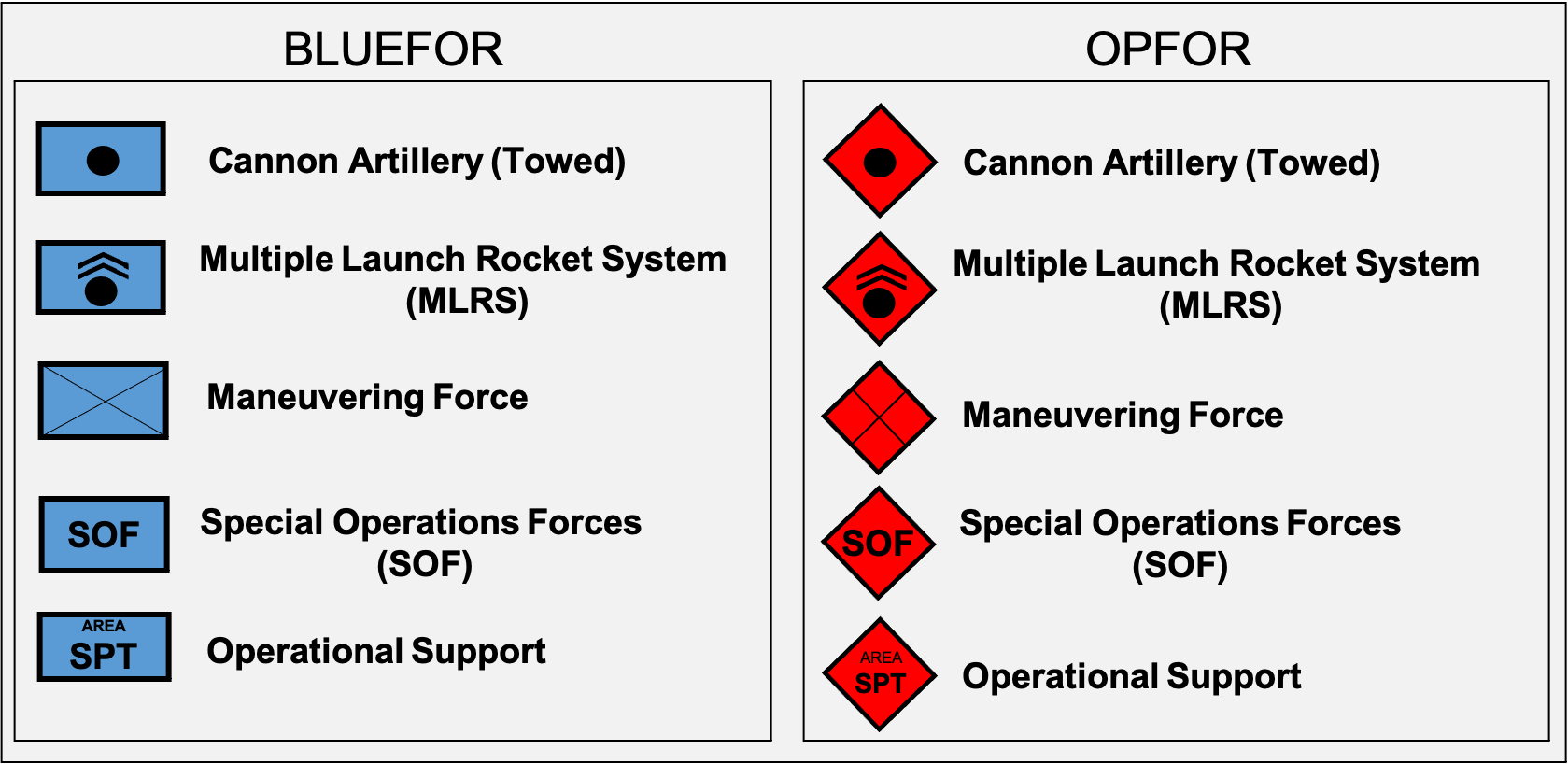}
    \caption{Assets and resources for the friendly (BLUEFOR, left) and opposition (OPFOR, right) forces. In the described MDO scenario it is assumed that all assets are  autonomy enabled formations (i.e., the assets contain autonomous robotic or software entities) for both BLUEFOR and OPFOR.}
    \label{fig:mdo_legends}
\end{figure}

\begin{figure}[t!]
    \centering
    \includegraphics[width=140mm,scale=0.5]{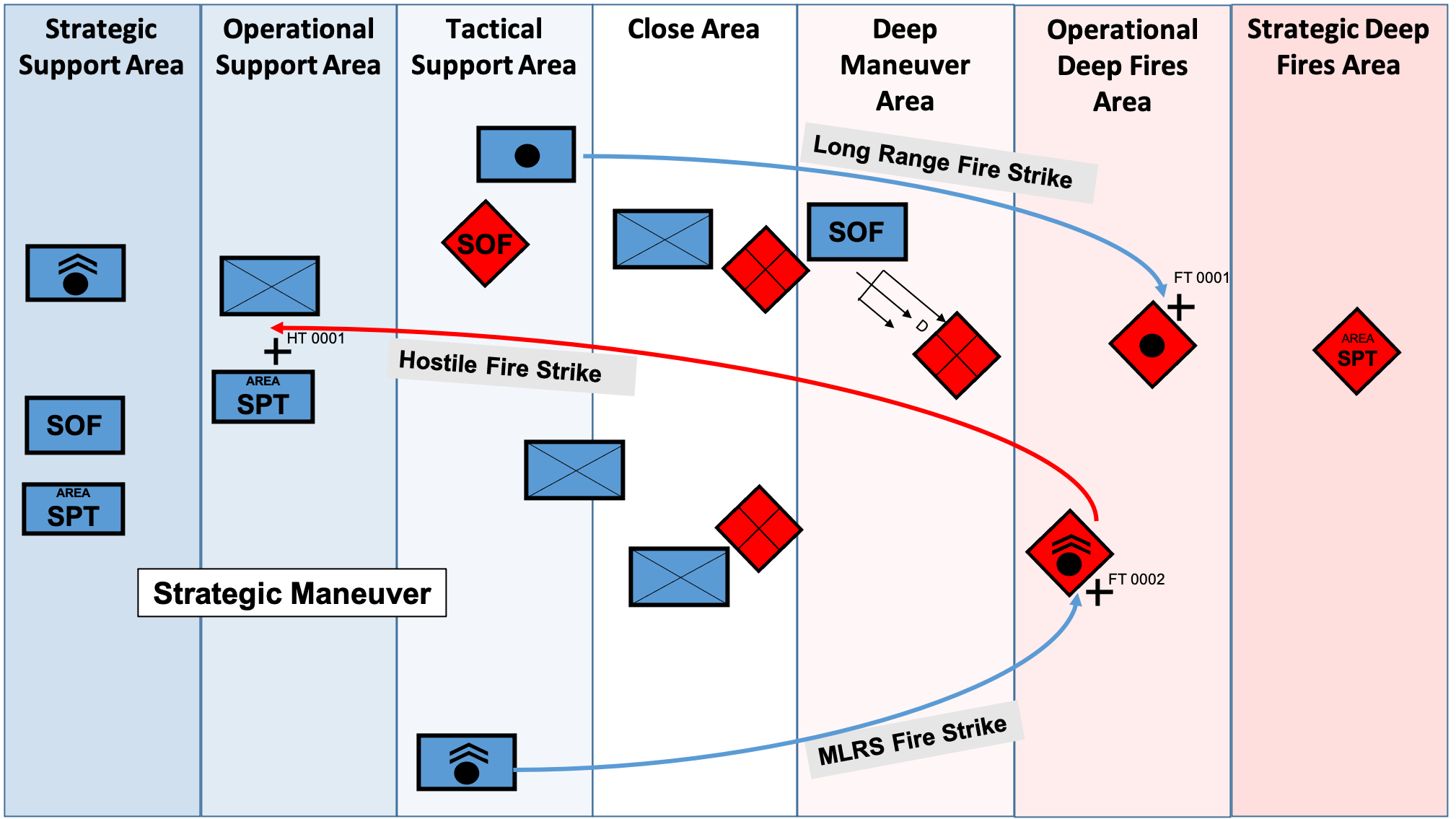}
    \caption{Adversarial forces (OPFOR) use long range missile and rocket fire to disrupt or destroy sustainment operations in the friendly (BLUEFOR) Strategic Support Area, which prevents friendly forces from engaging enemy maneuver elements in the Close Area on favorable terms. To counter this strategy, BLUEFOR conducts counterfire missions to destroy OPFOR long range fire systems located in the Deep Fires Area (blue arrow; Target Number FT0002). The three pronged arrow emanating from the BLUEFOR SOF in the Deep Maneuver Area represents a 'Disrupt' tactic which breaks up the adversary's formation and tempo.~\cite{adp102}}
    \label{fig:strat_man}
\end{figure}

In this section, an adversarial engagement scenario is provided that is centered on the use of selected long-range assets that inherently disrupt friendly force engagement. A legend is shown in Figure~\ref{fig:mdo_legends} to describe the  military symbology for selected assets and forces associated with the described MDO scenario. According to MDO~\cite{mdo} doctrine, in an armed conflict, adversarial long-range anti-access and area-denial (A2AD) fire systems (e.g., short-range ballistic missiles, long-range surface to air missiles and multiple launch rocket systems, which are protected from areal attack and hidden in unknown locations) can be used to deny friendly forces freedom of maneuver in a theater of operations. This is accomplished by combining intelligence, surveillance, and reconnaissance (ISR) assets with both lethal and non-lethal fires to attack friendly command structures, sustainment capabilities, and troop formations in the strategic and operational support areas (see Figure \ref{fig:mdo_legends}, OPFOR). These areas are the traditional staging ground for assets (e.g., troops and equipment) operating in the Close Area (see Figure \ref{fig:strat_man}). The adversary's ability to identify and engage targets well behind friendly lines (see Figure \ref{fig:strat_man}, Operational Support Area), causes those entities to be geographically separated from the Tactical Support and Close Areas, which effectively raises the attrition rate of friendly forces, referred to as \textit{stand-off}. Because the forward force is separated from strategic and operational maneuver support, adversarial  forces are able to take advantage of this friendly force isolation, and destroy them.

\begin{figure}[t!]
    \centering
    \includegraphics[width=140mm,scale=0.5]{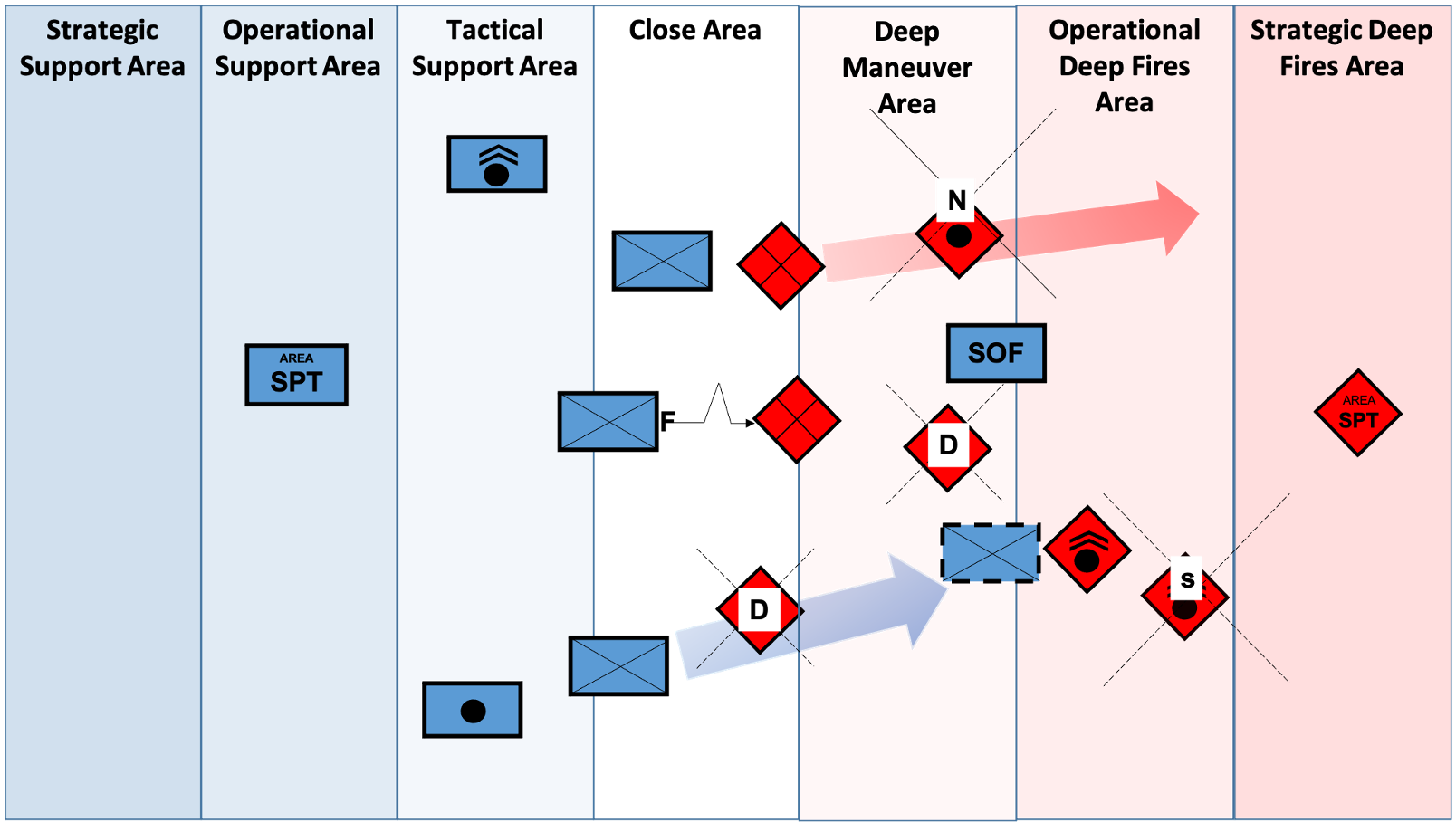}
    \caption{Suppressing (\textbf{S}) or neutralizing (\textbf{N}) the enemy long range fire systems and ISR assets enables friendly forces to penetrate the adversary's A2AD umbrella. This allows friendly forces to defeat the enemy in the Close Area and gives the maneuver commander the ability to exploit their success by rapidly moving forces into the Deep Maneuver Area to destroy (\textbf{D}) vulnerable enemy assets and pursue retreating enemy forces. The \textbf{F} indicates 'Fix' which effectively slows the adversary's movement. The thick arrows represent the direction of troop movements.~\cite{adp102}}
    \label{fig:disruption}
\end{figure}

The MDO~\cite{mdo} doctrine lays down a plan to defeat adversarial A2AD capability (i.e., \textit{stand-off}) so that strategic and operational maneuvers can enable forward deployed friendly forces to engage the adversary on favorable terms (i.e., penetrate and disintegrate A2AD systems to exploit freedom of maneuver). Here we focus only on the penetration and disintegration portions of an engagement\footnote{There are six joint functions from the doctrinal phases of a joint operation. These functions are: Command and Control (C2), Intelligence, Fires, Movement and Maneuver, Protection, and Sustainment. Penetration and disintegration are part of the 'Fires' and 'Movement and Maneuver' joint functions.} with adversarial A2AD systems by friendly (BLUEFOR) field army and corps that may entail the use of autonomous MAS in future battles. Further, it is speculated that all of the symbols shown in Figure~\ref{fig:mdo_legends} for both friendly (BLUEFOR) and adversary (OPFOR) forces will contain autonomy enabled formations (e.g., Robotic Combat Vehicles (RCV), automatic targeting systems, ground and aerial robotic ISR assets). Scenario diagrammatics for strategic maneuver and disruption utilizing this symbology with the autonomy enabled formations are shown respectively in Figure~\ref{fig:strat_man} and Figure~\ref{fig:disruption}. 

The adversarial A2AD fire systems create \textit{stand-off} by attacking the strategic and operational support areas as shown in Figure~\ref{fig:strat_man}. The friendly fires and air defense forces receive targeted intelligence from space and high-altitude surveillance (not shown) to strike high-value targets (i.e., MLRS) within narrow time windows to reduce adversarial position adjustments. In addition to surveillance, strategic \textit{stimulation-see-strike} can be employed to penetrate and disintegrate adversarial long-range fire systems~\cite{mdo}.

MARL can be used to strategically illuminate and track the locations of adversarial targets by exploiting adversarial doctrine and local observations from adversarial actions. Further, MARL trained autonomy enabled formations with a combination of highly mobile and distributed air and ground fires can begin to overwhelm adversarial long-range air defenses. Friendly forces may utilize MARL approaches trained to exploit adversarial TTPs for strategic maneuver of air-defense and ground fires. These autonomy enabled formations choose geographic locations based on surveillance data collected from strategic air-based stimulation. As adversarial long-range fire systems are neutralized, strategic and operational support units are able to advance (maneuver) towards the forward OPFOR (see Figure \ref{fig:strat_man}).

Adversarial forces identify friendly assets in the Operational Support Area using ISR assets and engage friendly forces with long-range fires systems (i.e., MLRS) from the Operational Deep Fires Area. These hostile fires disrupt the friendly force's ability to conduct traditional support operations in that area, which in turn, causes such activities to take place further back from the forward line of troops (FLOT). This creates geographical \textit{stand-off} by extending the battlefield, and straining supply lines. Further, this permits the hostile maneuver forces to engage friendly forces in the Close Area on terms favorable to the enemy \textit{fait accompli}. \footnote{fait accompli is a term that is used here to describe an event has already been decided before those affected hear about it, leaving them with no option but to accept it.} According to MDO doctrine, to eliminate \textit{standoff}, friendly artillery systems must identify, engage, and destroy hostile fires and ISR assets before they can be deployed. Friendly SOF assist this effort by disrupting supply and command and control (C2) nodes and providing targeting data for Joint Fires. This creates gaps in the enemy A2AD umbrella which can be exploited by maneuver commanders. \textbf{Under this coverage, friendly maneuver penetrates and then exploits gaps in the Close and Deep Maneuver Areas}.

Strategic formations of joint forces in the Close and Deep Areas, starting from the Operational Area, may be autonomous enabled formations (i.e., MAS) utilizing MARL trained policies to exploit adversarial TTPs (from doctrine), local observations, and ISR gathered information. Joint forces will coordinate between their ISR and long range precision fires capabilities to provide support for the forward deployed BLUEFOR forces as shown in Figure~\ref{fig:strat_man}. With support from strategic and operational units, forward forces with autonomous enabled formations can coordinate in Close and Deep Areas to isolate and defeat adversarial assets. This leads to the elimination of the adversarial forward maneuvering forces, (OPFOR) leaves long range fire systems vulnerable to ground attack (disintegration) in Figure \ref{fig:strat_man}).     

Joint Fires (i.e., friendly forces or BLUEFOR) suppress or neutralize adversarial long range fire systems, allowing friendly Maneuver Forces to move-in and defeat OPFOR in the Close Area (see Figure \ref{fig:disruption}). Friendly Maneuver Forces then exploit this advantage by destroying adversarial enablers in the Deep Maneuver Area (see \textbf{D} in Figure \ref{fig:disruption}). This causes the remaining adversarial maneuver formations to withdraw from the Close Area and establish a new front in the Deep Maneuver Area. This process repeats until strategic objectives are met or OPFOR is defeated. These coordinated activities could in theory be achieved in a collaboration between human Soldiers and an autonomous multi-agent system. Further, given that there is active research in the development and deployment of such autonomous systems, it is expected that battlefields of the future will need to consider scenarios like this for planning of strategic maneuver and disruption. 

This section has provided a scenario where autonomous enabled formations that may be trained with MARL approaches, however, the specific RL approaches to perform in such complex MDO environments have not been tested or may not yet exist.

\section{Insights and Conclusions}

Adversarial actors are becoming more advanced due to a number of factors, including scientific and technological progress made through proxy conflicts that test novel technologies. Coordinated strategic maneuver and disruption can be used by defense forces to give certain advantages over an adversary in future MAS autonomous warfare. In this article, some of the most prominent RL algorithms were discussed to uncover viable candidates for training MAS that can effectively perform strategic maneuver and disruption towards opening windows of opportunity in potential future military operations. A taxonomy of RL approaches was described with an  overview for the most prominent RL algorithms. 

It was found that most RL algorithms lack the capability of handling the complexities associated with potential future conflicts due to differences in training and test factors. However, two approaches appeared to utilize a combination of techniques (i.e., Dyna architecture and simulation-based search), which possess the potential to overcome both the inherent complexities associated with strategic maneuver and disruption in MDO, and the ever-changing dynamics of a battlefield scenario.

Since military engagement scenarios are highly dynamic and complex, it is no surprise that all current RL techniques fail to meet one or more challenges associated with these scenarios, and are not mature enough to be deployed off the shelf. A table is provided to summarize the capabilities and limitations of the RL algorithms described in this paper (see Table~\ref{tab:all_algos}. It is important to note that none of these approaches provide guarantees on performance when the testing environment is different in any way from the training environment, resulting in brittle or narrow solutions in the military engagement domain. Specifically, agents will typically need to be retrained if any changes are applied to the state or actions spaces after training has completed (e.g., adversary behavior or capabilities, own capabilities, team composition,  strategic team positioning, or mission objectives). Further, it is difficult to conceive of an RL technique that can handle the common concept of reinforcements to the battle if not explicitly trained for their inclusion. In addition, no RL technique has verified the ability to function properly for much longer durations than trained for (e.g., trained for a 10 hour engagement but must operate for days or weeks in deployment to overcome an adversary in practice). Finally, the algorithms described in this paper are not suited to perform multiple or  changing objectives, which is critical for strategic maneuver and disruption in military engagement scenarios.

\begin{table}[]
\scriptsize
\begin{tabular}{|l|l|l|l|l|l|l|l|}
\hline
Algorithm & Algorithm Type & \begin{tabular}[c]{@{}l@{}}Action-Space\\ Type\end{tabular} & Model Type & \begin{tabular}[c]{@{}l@{}}Scalable \\to Number\\ of Agents\end{tabular} & \begin{tabular}[c]{@{}l@{}}Change Number\\ of  Agents \\ after Training\end{tabular} & \begin{tabular}[c]{@{}l@{}}Multi-Agent \\ Reward Signal\end{tabular} & \begin{tabular}[c]{@{}l@{}}Algorithmic \\ Requirement for \\ Heterogeneous Agents\end{tabular} \\ \hline
PG & Single Agent & Continuous & Model-free & Unverified & No & N/A & \begin{tabular}[c]{@{}l@{}}Multiple instantiations\\  of the algorithm\end{tabular} \\ \hline
DQN & Single Agent & Discrete & Model-free & Unverified & No & N/A & \begin{tabular}[c]{@{}l@{}}Multiple instantiations \\ of the algorithm\end{tabular} \\ \hline
TRPO & Single Agent & Continuous & Model-free & Unverified & No & N/A & \begin{tabular}[c]{@{}l@{}}Multiple instantiations\\  of the algorithm\end{tabular} \\ \hline
DDPG & Single Agent & Continuous & Model-free & Unverified & No & N/A & \begin{tabular}[c]{@{}l@{}}Multiple instantiations \\ of the algorithm\end{tabular} \\ \hline
MADDPG & Multi-Agent & Continuous & Model-free & 10 & No & Individual & \begin{tabular}[c]{@{}l@{}}Single instantiation\\  of the algorithm\end{tabular} \\ \hline
MF-Q & Multi-Agent & Discrete & Model-free & 10 & No & Global & \begin{tabular}[c]{@{}l@{}}Single instantiation \\ of the algorithm\end{tabular} \\ \hline
M3DDPG & Multi-Agent & Continuous & Model-free & 10 & No & Individual & \begin{tabular}[c]{@{}l@{}}Single instantiation\\  of the algorithm\end{tabular} \\ \hline
EPC & Multi-Agent & Continuous & Model-free & 40 & Yes & Individual & \begin{tabular}[c]{@{}l@{}}Single instantiation \\ of the algorithm\end{tabular} \\ \hline
QMIX & Multi-Agent & Discrete & Model-free & 8 & No & Global & \begin{tabular}[c]{@{}l@{}}Single instantiation \\ of the algorithm\end{tabular} \\ \hline
RODE & Multi-Agent & Discrete & Model-free & 27 & No & Global & \begin{tabular}[c]{@{}l@{}}Single instantiation \\ of the algorithm\end{tabular} \\ \hline
\begin{tabular}[c]{@{}l@{}}Multi-agent\\ transformer\\ soft double \\Q-learning\end{tabular} & Multi-Agent & Discrete & Model-free & 1000 & Yes & Global & \begin{tabular}[c]{@{}l@{}}Multiple instantiations \\ of the algorithm\end{tabular} \\ \hline
MCTS & Single Agent & \begin{tabular}[c]{@{}l@{}}Discrete and \\ Continuous\end{tabular} & Model-based & Unverified & No & N/A & \begin{tabular}[c]{@{}l@{}}Multiple instantiations \\ of the algorithm\end{tabular} \\ \hline
MAMCTS & Multi-Agent & \begin{tabular}[c]{@{}l@{}}Discrete and \\ Continuous\end{tabular} & Model-based & 1000 & Yes & Individual & \begin{tabular}[c]{@{}l@{}}Single instantiation \\ of the algorithm\end{tabular} \\ \hline
Dyna-Q/Q+ & Single Agent & \begin{tabular}[c]{@{}l@{}}Discrete and \\ Continuous\end{tabular} & Both & Unverified & No & N/A & \begin{tabular}[c]{@{}l@{}}Multiple instantiations \\ of the algorithm\end{tabular} \\ \hline
\end{tabular}
\caption{Summary for findings from algorithmic review of recent literature.}
\label{tab:all_algos}
\end{table}

There are a three algorithms (EPC, MAMCTS, and Dyna-Q/Q+) that stand out as potentially viable options for training a MAS for strategic maneuver and disruption in MDO. Although these algorithms stand out, they all have their drawbacks that would cause them to fail under certain conditions in military engagement scenarios.

Among the list of algorithms shown in Table~\ref{tab:all_algos}, EPC is one with the potential to perform decently in some portions of the military MDO scenario shown in Figures~\ref{fig:strat_man},~\ref{fig:disruption}. EPC utilizes an individual reward, can be scaled up to 40 agents (indicating some level of handling reinforcements or degraded forces), and operates in a continuous space (desired for small-scale rapid actions), which implies it can operate in tactical level engagements where a variable number of heterogeneous agents must coordinate actions to defeat a dynamic adversary. However, given that EPC is model-free, the state and action spaces may be quite large, making training potentially untenable for  complex or multi-domain tasks. 

In contrast to EPC, MAMCTS has been shown to scale up to 1000 agents (after training), operates with individual rewards as well (indicating support for heterogeneous agents and different objectives), functions in both continuous and discrete environments (allows for both tactical and operational or strategic level engagements), but is strictly model-based (requires a model of the environment which may be unavailable), which may lead to substantial difficulties for practical implementation in military engagement scenarios. 

Although Dyna-Q/Q+ has been unverified in scalability (unlike EPC and MAMCTS), is inherently a single agent algorithm (indicating the need for multiple algorithmic instantiations for multi-agent implementation), and may not support variations in agents (to accommodate reinforcements or lost forces), it can operate in continuous and discrete domains (for both tactical and operational or strategic level engagements), and most importantly, utilizes both model-free and model-based approaches, which allows for a basic model of the environment to facilitate exploration, unlike the drawbacks of strictly model-free or model-based approaches alone. The strength of  Dyna over EPC and MAMCTS is that it can expand upon a simple model of the environment (model-based) with state and action space exploration (model-free) to encounter novel tactics and strategies from simple know doctrine based maneuver. Further, a multi-agent extension of Dyna could permit this technique to match or overcome the strengths of both EPC and MAMCTS, which might be achieved with a centralized training signal (as is done with the majority of MARL algorithms). Therefore, the combination of model-free and model-based approaches gives algorithms like Dyna, an advantage over other algorithms for application in coordinated MAS military engagement scenarios. As a specific example, if we provide an algorithm a model (model-based) of how to engage an adversary (e.g., maneuver formation for tactical engagement), and then allow the algorithm to explore the state and action spaces (model-free), then the result may be a more effective engagement approach that minimizes friendly force losses and maximizes adversarial casualties. 

In general, for a military engagement scenario we need a RL algorithm that is robust: can maintain the training performance with changes in the testing environment and does not need to be retrained if new capabilities are be added, scalable: capable of training a large number of heterogeneous agents which can be reinforced during testing and capable of performing for a long operational duration. We think the future direction of the military should be focused on integrating the strengths of model-free and model-based algorithms for coordinated MAS to execute strategic maneuver and disruption. Alternatively, if we cannot use model-based algorithms (due to the lack of a model) there needs to be a focus on scalable model-free algorithms,  where centralized training approaches (i.e., centralized training with decentralized execution - CTDE) can be a computational bottleneck but essential for optimality (i.e., best performance given a reward structure). To remove the bottleneck of centralized training, finding partially independent sub-tasks in the state space and a hybrid training approach (between centralized and decentralized) might be an alternate potential future direction for model-free techniques.

Further, additional investigation is needed to illuminate military strategy that facilitates the utilization of MAS in engagement scenarios. In obvious cases, it is desirable to send fully autonomous MAS into high risk situations (i.e., where expected causality rates are high), however, it is insufficient to simply expect that a MAS will be capable of achieving the Mission in the absence of human oversight or intervention due to current technological limitations. Therefore, in future work, it is expected that a robust set of engagement scenarios will be identified. Finally, this line of work will lead to the eventual integration of semi/fully-autonomous MAS for coordinated strategic maneuver and disruption where possible.

\section{Acknowledgements}

This project was supported in part by an appointment to the Science Education Programs at National Institutes of Health (NIH), administered by ORAU through the U.S. Department of Energy Oak Ridge Institute for Science and Education. Further, this work was supported by the US DEVCOM ARL ERPs, which have provided the resources to describe and pursue solutions towards operationalizing science for strategic maneuver and disruption with MAS in MDO. ARL's ERPs provide a programmatic path for developing and implementing intelligent MAS. Given that Army research programs provide US defense operations with answers to critical research questions, these ERPs specifically enable research that can provide a path towards coordinated autonomous MAS that can overcome the challenges associated with, 1) the complexity of an environment, 2) adversary tactics and capabilities, 3) own capabilities (i.e., gain new capabilities, lose previous capabilities, or have capabilities altered), 4) team composition (e.g., adding, removing, or swapping of team mates), 5) strategic team positioning, entry, navigate (maneuver) to support forces and overwhelm an adversary, and 6) mission objectives.

\bibliographystyle{SageV}
\bibliography{references}

\end{document}